# An Exploratory Study of Forces and Frictions affecting Large-Scale Model-Driven Development[*]

Adrian Kuhn, Gail C. Murphy, and C. Albert Thompson

Department of Computer Science
University of British Columbia, Canada

**Abstract.** In this paper, we investigate model-driven engineering, reporting on an exploratory case-study conducted at a large automotive company. The study consisted of interviews with 20 engineers and managers working in different roles. We found that, in the context of a large organization, contextual forces dominate the cognitive issues of using model-driven technology. The four forces we identified that are likely independent of the particular abstractions chosen as the basis of software development are the need for diffing in software product lines, the needs for problem-specific languages and types, the need for live modeling in exploratory activities, and the need for point-to-point traceability between artifacts. We also identified triggers of accidental complexity, which we refer to as points of friction introduced by languages and tools. Examples of the friction points identified are insufficient support for model diffing, point-to-point traceability, and model changes at runtime.

## 1 Introduction

Model-driven engineering (MDE) is the primary use of, often visual, models for software engineering. Although technical approaches of model-driven engineering are well-documented, there is a paucity of information about how humans interact with and adapt to the technology.

In this paper, we investigate the human aspects, reporting on an exploratory qualitative study conducted at General Motors, a large automotive company who makes extensive use of model-driven engineering.

Our study involved interviews with 20 engineers and managers. These interviews took an individual-out perspective, that is from the perspective of engineers to their context, focusing on how an individual is applying and grappling with model-driven technology to complete assigned goals. We analyzed the interviews to identify triggers of complexity that may arise when working with software models and how those triggers compare to those found in more traditional forms of source-based development.

We look at triggers of complexity in terms of forces and points of friction. The forces are likely independent of the particular abstractions chosen as the basis of software development and thus should be considered in the design of any new abstractions. Our notion of forces is similar to Brooks's notion of essential complexity from his "No Silver Bullet" essay [1]; they transcend the modeling technologies used. Related to each force we also identified points of friction, which are akin to Brooks's notion of accidental complexity; namely complexity introduced by languages and tools.

---

[*] To appear in proceedings of *Model Driven Engineering Languages and Systems, 15th International Conference (MODELS 2012),* LNCS Springer, 2012.

Through our study, we identified four forces and five points of friction that affect the use of model-driven engineering at the industrial site we studied, which may provide insight into model-driven engineering in general. They are as follows:

– Teams are typically working on multiple versions of the same software model (force), yet engineers lack proper tooling to identify and share diffs (friction).
– Domain experts use a rich set of visual and formal languages to invent novel algorithms (force), yet they lack tool support to define their own little visual languages (friction) or pluggable ad-hoc type systems (friction).
– Inventing novel algorithms for vehicle control is an exploratory activity (force), while the needs of early prototyping are well addressed by in-silico simulations, testing on actual vehicles, which occurs later in the process, suffers from lacking tool support for model changes at runtime (friction).
– Requirements documents and software models need be kept consistent across development iterations (force), yet engineers lack proper tooling to track point-to-point correspondences between corresponding artifacts (friction).

We believe that the forces and frictions we have identified through this empirical study can help software engineering researchers understand the context in which model-driven software engineering occurs in practice and that the friction points we identified can influence new modeling languages and tools. The specific results of this study can also help those adopting model-driven engineering to understand cognitive issues that may impact the use of MDE.

This paper makes four contributions:

– it introduces the notion of forces and points of friction in tooling to describe the impact of technical issues in the use of model-driven engineering,
– it identifies and presents four forces that may significantly impact the use of model-driven engineering,
– it identifies and presents five points of friction in existing language and tool support for model-driven engineering,
– it provides points of comparison with source code development to help tease apart essential and accidental complexity.

The remainder of this paper is structured as follows. Section 2 discusses methodology of our field study, Section 3 presents the software development process at the organization we studied, Section 4 presents the findings of our study, enumerated as contextual forces and points of frictions, Section 5 discusses our findings in the general context of model-based design, Section 6 presents related work, and Section 7 concludes with concluding remarks.

## 2  Methodology

To enable the gathering of detailed, rich and contextual information about model-driven engineering, we chose a qualitative study approach. We visited the industry of interest (General Motors) on two separate occasions, collecting data constructed through semi-structured in-depth interviewing. We interviewed 12 engineers and 8 managers. Overall, the engineers we interviewed came from four different teams from different company departments. All teams were global, that is spread across sites in India and America, however we interviewed people from the American sites only. The 12 engineers selected for interviews were sampled from several roles however their profiles are similar, that is they all work with the same process and use the same modeling technology. Each interview was 90–120 minutes long, recorded on tape and transcribed for encoding by one of the authors of this paper.

In a first visit, we interviewed 10 participants from both management and technical roles to familiarize ourselves with the software process used in the automotive industry. Based on what we learned from the first interviews, in our second visit, we interviewed an additional 10 participants, all of them working with software models but in different roles. The interviews were semi-structured, following an exploratory case-study approach where open ended questions are asked in order to identify research hypothesis for future studies [2]. We asked participants to describe their work, how their work fits into the process of the organization, with whom they interact on a weekly basis, and which artifacts are the input and which are the output of their work. We also asked to see current or recent examples of artifacts on which they were working.

We transcribed the 12 interviews with engineers (4 from the first visit and 8 from the second visit). We encoded the transcripts and from this encoding, we distilled the contextual forces and points of friction presented in this paper. We encoded the interviews by tagging sentences with hashtags as if they were tweets. We then used a series of tag clouds to identify patterns in the data, merging and splitting tags as we saw need. We did two passes over the tags, a first one to identify all forces and frictions that shape the work of the participants, and a second pass to identify forces and frictions that might provide the basis for general hypotheses on model-driven engineering, ruling out those that are specific to the organization under study.

The data presented in this paper is largely from the in-depth interviews with 12 engineers. These engineers worked in the following roles: 2 domain experts, 7 software modelers, and 3 testing engineers. The participants had an average of $12.5 \pm 8$ years of professional experience with software engineering and an average of $4.5 \pm 4$ years of professional experience with modeling; their backgrounds were electronic engineering (9 mentions), mechanical engineering (4 mentions), computer engineering (3 mentions), and software engineering (1 mention). There are more than 12 mentions as some engineers had two degrees.

*Threats to Validity:* We selected all participants from the same organization, whose common context and corporate culture may bias the results. We were for-tunate however to interview participants from four global teams and a wide variety of roles, providing us with multiple views on the cognitive issues of working with modeling technology. Although a study at one organization is not sufficient to make broad generalizations, this initial data can provide at least one practical reference point of context that is otherwise often absent in language and tool design. This practical reference point can provide a basis for more specific hypotheses to study in future empirical work in this area.

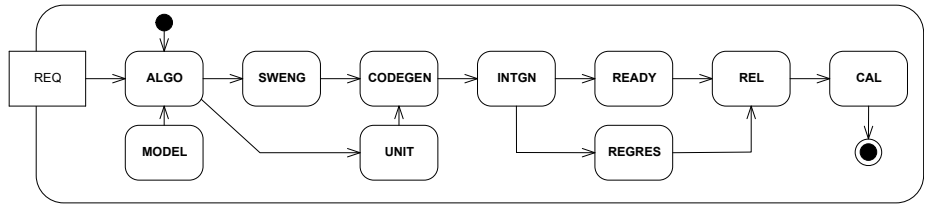

**Fig. 1.** Software development process: the doodle shows all stages of an iterative six week release. From left to right, the stages of the process are: REQ) requirements collection, ALGO) algorithm design, MODEL) in-silico simulations; SWENG) software model development; CODEGEN) code-generation; UNIT) unit testing; INTGN) integration on embedded chips; READY) readiness testing; REGRES) regression testing; REL) internal release; CAL) calibration on actual vehicles.

## 3 Modeling at General Motors

To enable interpretation of our qualitative study results, we provide an overview of the software development process at General Motors. We begin by describing the overall software development process used, followed by a more in-depth description of the various roles involved with software development and the artifacts produced and consumed during the process.

### 3.1 Process Overview

Figure 1 shows the software development process commonly used in the automotive industry. While the figure depicts a sequential flow from requirements to deployed software on the vehicle, the actual process happens in iterative releases of six weeks with different stages of the process running in parallel on subsequent releases. Development begins with requirements collection, which typically happens outside the software development team (REQ in Figure 1). The requirements are consumed by domain experts of the team who perform algorithm design (ALGO), which includes running tests of developed models on in-silico simulated vehicles. Software modelers consume developed algorithms (either requirements or model patch) to produce software models from which code can be generated in an automated step (CODEGEN); code generation is 100% automated, a special team of "language designers" maintains the rules used for code generation. Test engineers use the results of algorithm design and generated code to perform unit tests; these engineers work primarily with source code. Integration engineers take care of integrating produced software to embedded chips. Test engineers take the results and perform readiness (READY) and regression tests (REGRESS). Every six weeks, teams downstream in the process receive new software that is calibrated on the car (CAL). This step involves calibrating the parameters of the typically generic features developed in the software to a specific car model.

Each team following this process typically owns a single feature and the models that describe that feature. The models for a single feature are reused for different versions (world region, national legislation, car model and year) of a particular car. As described above, special teams do exist that provide the other teams with infrastructure and code-generation rules.

### 3.2 Roles

A software development team responsible for a feature consists of about a dozen people working in different roles and possibly different countries. Through our interviews we learned about four different roles.

**Domain Experts** are responsible for maintaining requirements documentation and inventing novel algorithms. In the former responsibility, domain experts work more distinctly from software modelers. In the latter responsibility, domain experts work closely with software modelers, including drafting changes to models on which the software modelers work. The algorithms that the domain experts are designing are not so much computational algorithms but rather involve the physics of a vehicle. Most domain experts thus have a strong background in electrical or mechanical engineering, but typically no formal education in software engineering.

**Software Modelers** implement and maintain models as specified by domain experts. Software modelers are responsible for three to four models and are in close collaboration with the domain experts who own the corresponding requirements documentation. Software modelers use the MatLab Simulink[1] or IBM Rhapsody[2] tools; we describe more about these tools in the next section. When the models compile, they are passed on to integration engineers for integration into a release. Most software modelers have a background in mechanical or electrical engineering, some have a minor in computer or software engineering, but this is the exception rather than a rule.

**Test Engineers** perform delta and regression testing of releases and are responsible for root cause analysis of an incoming anomaly report (i.e., bug reports). Test engineers typically work with generated sources rather than models. Test engineers are exposed to all artifacts in the process and tend to have the broadest knowledge of a team's feature. New hires are often first assigned a test engineering role before moving on to a software engineering role. The professional background of test engineers is the same as for software modelers.

**Code-generation engineers** belong to a special team that owns and maintains the rules used to automatically generate source code from the software models. These experts also publish modeling guidelines and naming conventions. Even though not formally established by the process, software modelers are often in close contact with code generation engineers, providing them with feedback and getting help when they struggle with code-generation issues.

---

[1] http://www.mathworks.com/products/simulink
[2] http://www.ibm.com/software/awdtools/rhapsody

### 3.3 Artifacts

Requirements and software models are the main representations used in the software development process. We describe these two artifact types and highlight four kind of secondary artifact types that are relevant to our results.

**Requirements Documents** are specifying features and owned by a team. These documents are maintained by the domain experts. The requirements documents that we saw are loosely structured MS Word documents, typically containing a mixture of natural language text, pseudo-code and figures. Figures within these documents often use problem-specific visual languages and are created manually. In maintenance teams, requirements documents are changed first and drive subsequent changes to software models. In innovative teams, domain experts explore the solution space by drafting changes to the software models themselves and requirements documents are updated once the algorithms stabilize.

**Software Models** are created by software modelers with either Matlab Simulink or IBM Rational Rhapsody. While the two are interchangeable in the process and used by the same roles, they are technically quite different:

- Matlab Simulink is a model-based design tool focused on the design of control flows. Simulink models are written in a low-level visual language which resembles the visuals of circuit diagrams. Code generation with Simulink is automated but cannot be customized.
- IBM Rhapsody is a model-driven engineering tool. The structure of Rhapsody models is specified using UML class diagrams, where engineers can choose between visual and non-visual representations, and behavior is specified using either blocks of C-code or state machine diagrams. Code generation with Rhapsody is automated and highly customizable.

Figure 2 shows a sketch of a typical Simulink model that implements part of a feature. A typical model consists of about 100,000 blocks and a dozen nested layers. From left to right, we see model layers of increasing nesting level: 1) the top most layer of the model, which is structured according to the modeling guidelines with "the function" on top and other diagnostics function on the bottom; 2) the second layer, zooming into the function, showing 96 input signals and 45 output signals; 3) One of many layers that serves to convert the unit and magnitude of input signals, typically each of those layers corresponds to a paragraph in the requirements document; 4) about a dozen layers deeper, program logic such as conditionals and loops are laid-out as a graphical circuit with each major block corresponding to a paragraph in the requirements document; 5) further down inside one of the blocks with program logic, basic arithmetic operations, such as addition and division of numbers, are modeled using the visual language of circuits rather than using mathematical notation.

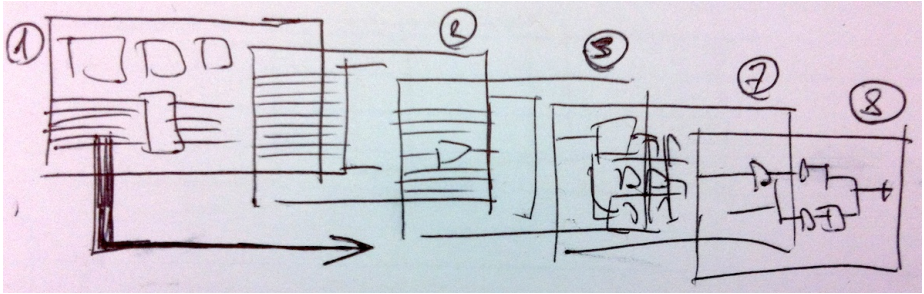

**Fig. 2.** Sketch of a Simulink model: from left to right we see model layers of increasing nesting level, starting with the entry function down to implementation logic.

| Sec | Friction | $P_1$ | $P_2$ | $P_3$ | $P_4$ | $P_5$ | $P_6$ | $P_7$ | $P_8$ | $P_9$ | $P_{10}$ | $P_{11}$ | $P_{12}$ |
|---|---|---|---|---|---|---|---|---|---|---|---|---|---|
| 4.1 | Insufficient model diffing | • | • | • | • | • | • | • | • | • | • | • | • |
| 4.2 | Need for Visual DSLs | | | | | | | | | | • | • | • |
| 4.2 | Hungarian notation as types | • | • | | | • | • | | | | | | |
| 4.3 | Need for exploratory programming | | | | • | | | | | | • | • | • |
| 4.4 | Lack of P2P traceability | • | • | • | • | • | • | • | • | • | • | | • |
| | **Role** | S | S | S | T* | S | S | T | T | S* | X* | X* | S* |

**Table 1.** List of observed frictions by participant numbers. Roles are X) domain expert, S) software modeler, and T) test engineer; an asterisk indicates that their team focused on inventing novel algorithms rather than on maintenance of stable technology.

**Auto-Generated Source Code** is often defined as a secondary artifact type; however, this code serves as the primary artifacts used by test engineers and sometimes, for diffing and change tracking, by software modelers.

**Code-Generation Rules** are used for automated code generation. These rules are maintained by a special team of code-generation engineers.

**Model Patches** are used by domain experts when they invent novel algorithms to exchange their model prototypes with the software modelers. These patches are not a formally defined part of the software development process and thus are *ad-hoc* artifacts. Model patches take many forms, such as Excel spreadsheets with annotated screenshots of a model.

**Tests** ensure that software models are implementing a feature as specified in the requirements documents. Within a team all tests are owned by the test engineers. Many tests require manual manipulation on a workbench, that is a partial vehicle in the testing lab, while other tests are fully automated and run in an in-silico simulation of the vehicle and its environment.

## 4 Results — Forces and Frictions

From our analysis of the interview data, we identified four forces and five points of friction. Using Brooks's terminology, forces are indicators of innate complexity while frictions introduce accidental complexity [1].

Table 1 documents the observed frictions by participant number. As we distilled the forces and frictions from the encoded interview data, we tried to identify those triggers of complexity that are not specific to the organization that we studied. We focused on those that are more general and thus might form the basis for more specific hypotheses in future research work.

### 4.1 Force: Need for Diffing in Software Product Lines

Engineers in a team are often working on different versions of the same artifact. Thus, engineers need to identify the changes in a model and, possibly, share those changes with engineers in other roles. Working with multiple versions is a result of business needs and thus a contextual force, independent of the primary abstraction used for software representation, i.e., models or code.

Internal releases happen every six weeks, however, the length of a full iteration might be longer. It is common for multiple engineers, in the same team, to work on different releases of the same model. In particular, we learned that domain experts typically use previous releases to prototype the changes that are supposed to drive future releases. Thus, domain experts need to exchange those prototypes as model patches with the software modelers. Also, test engineers reported that they need to learn about the most recent changes to a model under test.

**Friction: Insufficient Support for Model Diffing** (12/12 interviews). Engineers use version control to keep track of different versions and revisions of the same model. However, they experience friction when merging and handling comparison of these versions.

Although there are commercially available third-party tools that offer diffing capabilities for modeling, the engineers we interviewed described that they are limited in their scalability and in their usability. Engineers described the experience of using these tools as *"going blind"* ($P_{10}$) and leading them to make mistakes. Engineers seem to prefer a linear reading path of textual diffing in order to make it easier for them to *"not miss a change"* ($P_9$). Model-based approaches which highlight the changes in the spatial and possibly nested visual representation do not provide that kind of linear reading path.

We learned about several different strategies that engineers use to work around the lack of model diffing:

– When coming back to their own work, or keeping track of changes for code reviews, software modelers adopted a habit of documenting all model changes with comments. One example is the unique identifier of the current work ticket used as a marker, such that a search for this marker returns all model changes. This approach is the same as the approach adopted to handle missing point-to-point traceability, which is discussed in Subsection 4.4.
– When comparing versions for regression testing and root cause analysis, test engineers use textual diffing tools on the auto-generated code, which puts them at risk to misinterpret the modeler's intention of a feature.

- When inventing algorithms and prototyping on their own branch of a model, domain experts often use screenshots to communicate their changes back to the software modeler who owns the model. They take screenshots where the changes before and after, marked them in red, and share them in a PowerPoint slide deck as an *ad-hoc* model patch which is either emailed or attached to a change ticket.

The engineering needs for model diffing are similar to those found in traditional source code development. We did not hear in the interviews that an increased level of abstraction in representation (that is modeling rather than code) leads to an increased need for semantic diffing. Our observations suggest that end users would like to have more scalable and usable syntactic diffing rather than semantic diffing. As reported by the engineers who are falling back to textual diffing of auto-generated source, syntactic diffing is in their words *"more than good enough"* ($P_7$) for most use cases; in particular when diffing is needed to track the changes from one version to another.

### 4.2 Force: Need for Problem-Specific Expressibility

Domain experts use a rich set of visual and formal languages to invent and design their algorithms. The requirements documents that we encountered in our study made use of a rich and diverse visual language to describe the desired behavior of algorithms. Some of these languages are by virtue of the domain expert's training as mechanical or electrical engineers, whereas others of those languages are a result of the domain expert's struggle to find the best way to explore the problem and solution spaces of their inventions.

We found that the modeling tools in our study, while providing the specialized team of code-generation engineers with powerful abstractions, do not empower end-users, that is domain experts and software modelers, to define their own problem-specific "little languages" [3]. We highlight two major points of friction related to insufficient expressibility that we identified: visual languages and *ad-hoc* types.

**Friction: Lack of Problem-Specific Visual "Little Languages"** (3/12 interviews). Domain experts often need to prototype their innovations in a software model, yet the visual language of modeling tools limits their ability to express themselves. The notations that domain expert use to talk and think about their algorithms are those found in mechanical and electrical engineering.

For example, a domain expert might be prototyping a novel clutch control. In the requirements documents, the domain experts might describe the behavior of two dependent variables as a graph with two signals in time, quote: *"there are pictures in here of how I want the data to behave, and when I am done I want to see this [on the oscilloscope] on a car."* ($P_{10}$) Yet, when the domain expert uses software models to explore the solution space of the novel algorithm, he has to constantly translate back and forth between his mental model and the programing constructs. The domain expert cannot just draw a graph of the expected behavior and have appropriate code generated.

Support for domain-specific modeling might alleviate this point of friction, please refer to the discussion in Subsection 5.3 for more information.

**Friction: Hungarian Notation Used as Ad-hoc Types** (4/12 interviews). We also found that some teams used Hungarian notation to denote physical unit and magnitude of signal names in the Simulink models. Hungarian notation was popular in software engineering before the introduction of type systems. It is a naming convention where variable names were prefixed with abbreviations indicating the type of a variable, e.g., `szName` for a variable storing a username as zero-terminated string.

The software engineers described to us how they use Hungarian notation to denote the physical type and magnitude of signals in their models. For example, they use a prefix to indicate that a signal is temperature in degree Celsius and that it is a fixed-point number with base 10 and radix 2. The printed list of all prefixes used in the system fills four pages and they keep them close to the keyboard, to have them always ready when working with the models. Engineers use these prefixes to make sure that values are properly converted and normalized before use. However, these coding conventions are only manually, not automatically, verified.

Support for problem-specific type systems in modeling technology, as for example pluggable types [4], might alleviate this point of friction.

### 4.3  Force: Inventing Novel Algorithms as an Exploratory Activity

Developing algorithms for vehicle control is an exploratory activity. While the needs of early prototyping are well addressed by in-silico simulations, this is not the case for later stages where novel algorithms are tested on actual vehicles. As engineers are testing software "on the car," during a test drive on the proving grounds, they often encounter the need for updates to the software system. This need has been reported by domain experts and software engineers in those teams who work on inventing novel algorithms.

**Friction: Long Build-Cycles Prevent Live Modeling** (4/12 interviews). The build process of the model-driven tool chain may take up to several hours. As a result, when working "on the car," as soon as the need for a software change arises the test drive has to be interrupted and rescheduled for another day. Engineers reported that build times with an older C-based tool chain had been in the half-hour range and thus within a tolerance interval where it had been possible to continue the test drive on the same day. Ideally though, when working "on the car" engineers should be able to apply model changes at runtime and continue their test drive instantly.

This point of friction might be alleviated by an abstraction which does away with compile-build-deploy cycles, such that changes to the software can be applied at runtime. Technologies that allow a form of hot-swapping of quickly-generated code from models might be a means to address exploratory adaptation of software at runtime. Such technologies would not be limited by the processing power of target hardware (embedded control units) since while working "on the car" those chips are stubbed by more powerful hardware anyway.

### 4.4 Force: Need for Traceability in Incremental Release Cycles

A major theme that appeared throughout the interviews was the need for traceability between specification documents and software artifacts. Requirements documents, software models and tests are all essentially different representations of the same information, which need be kept consistent as those artifacts are independently updated with each release cycle.

While the content management system used in the organization provides engineers with document-to-document traceability, for many tasks point-to-point traceability is required. Engineers need to be able to quickly navigate from a visual block in the software model to the corresponding paragraph in the requirements document, or the corresponding test, or even the auto-generated sources, and vice versa. While this need is essentially representation independent, the introduction of software models as an additional layer of abstraction exponentially increases the traceability needs of engineers.

**Friction: Lack of Point-to-Point Traceability** (12/12 interviews). Currently, engineers establish traceability by relying on naming conventions. All interviewed engineers mentioned the use of markers as a work-around for missing point-to-point traceability. We found that engineers have adopted a habit of using change ticket identifiers as markers to establish point-to-point traceability through manual search. This is similar to one of the habits adopted to tracking changes between model versions as discussed in Subsection 4.1.

While this workaround establishes limited point-to-point traceability, the approach is inefficient and fragile. If engineers forget to mark one of the documents with the unique identifier, traceability is broken. In addition, while names contained in software artifacts are verified by code generation or compilation, names contained in specification documents often contain spelling errors or use old names that predate a renaming refactoring. Spelling differences make it hard, if not impossible, to navigate these traceability links using keyword search.

## 5 Discussion

In this section, we discuss our observations in the context of model-driven engineering and provide points of comparison with source code development to help tear apart essential and accidental complexity.

### 5.1 On the Terminology of "Model"

In our interviews, we found that the terms "model" and "modeling" were used ambiguously. Engineers generally did not refer to their work as "modeling" but used the terms "auto-coding" and "hand-coding." These terms were used to differentiate between working with tools which include a step of code generation versus writing C-level code manually. Engineers used the term "model" ambiguously to refer to software models, as well as the *plant models* used for the in-silico simulation of vehicles. Engineers also used the term "simulation" ambiguously to refer to running the in-silico simulation of the plant models, as well as to running software models from within the modeling tools as opposed to running the auto-generated sources.

We believe the terminology we observed is mixing model-based design (MBD, an approach in system engineering for disentangling the development of control software and corresponding vehicles, using in-silico modeling while vehicles are not yet available) and model-driven engineering (MDE). The ambiguous use of terminology can be explained if we look at model-driven engineering as a division of labour between a few specialized language designers and many modelers. After all, the software engineers do not have to understand the full complexity of modeling, this is up to the specialized code-generation engineers. However, we found that points of friction in modeling tools, in particular the insufficient support of model diffing, may break the abstraction and nevertheless expose engineers to these complexities.

### 5.2 On Visual Models and Linear Reading Paths

During our interviews we learned about heated controversy around modeling among engineers, and whether hand-coding is superior to code generation. While some of the critique was targeted at the long build cycles of the modeling toolchain (see Subsection 4.3), much of it was concerned with the visual representation of models and its lack of abstraction such as scopes and subroutines.

Without the linear order of text lines, which is superimposed upon source code, visual programming as found in models has no linear reading path and can possibly stretch in all directions, left, right, top, bottom, and even down to the next nesting level. While modeling guidelines try to alleviate this by imposing a flow from top-left to bottom-right, engineers struggle with reading visual models as to make sure they are not missing a part of their work to-be-done. Engineers expressed difficulties with reading order both when navigating (see Subsection 4.4) and changing (see Subsection 4.1) models.

For example, when doing readiness testing, all changes in the current release need be covered with tests and no single change must be missed. One participant gave an account of a case where they printed a whole model, put all layers up on a huge wall and worked together on the wall-sized printout to make sure they *"can walk through the complete model and don't miss a block"* ($P_7$).

Another engineer showed us how she uses a numbering scheme to reduce the spatial complexity of her visual models down to linear reading path, quote: *"this is just [a] little help for myself, we don't have to do this, I add numbers to each blocks, like 7 and 8 and 9, and then 8.1 and 8.2 and deeper down 8.3.1.6, so I can read the model from top to bottom."* ($P_{12}$) The same motivation, that is introducing linear reading paths, was brought forward by engineers when they described their practices of sharing model patches as PowerPoint decks and or when motivating their preference of textual diffing tools.

Related to this point, when offered an alternative to visual programming engineers seem to prefer non-visual representations. The Rhapsody tool offers engineers an alternative to visual programming which is editing the class diagrams through a tree view and property dialogs. Engineers seemed to prefer this option over visual modeling of UML class diagrams and they even reported that, to their best knowledge, the visual representation of class diagrams is not used by other engineers either.

### 5.3 On Problem-Specific Needs of Modelers

While model-driven engineering at GM provides the specialized team of code-generation engineers with powerful abstraction to capture domain-specific architectures, it does not empower its end-users, i.e., domain experts and software engineers, to express their own problem-specific languages and type systems.

In general, the visual language of domain experts seems to be much richer and broader than the languages provided by modeling tools. In particular, there seems to be a need for problem-specific "little languages" that can be defined on the fly. Currently, domain experts are unable to create new abstractions that would allow them to achieve productivity gains in algorithm innovation. Neither Simulink, which is largely a visual representation of common coding patterns, nor Rhapsody, by virtue of its limitation to the UML standard's visual languages, offer the ability to define the kind of rich visual languages that we learned about from the domain expert's requirements documents.

Visual programming in Simulink traces its ancestry to circuit diagrams and aims at expressing low-level programming constructs such as conditionals and mathematical operators with the visual language of circuit diagrams. Mathematical operations and conditionals are each represented as single blocks. While this language is visual, it does not seem be an actual abstraction from source code. Even worse, as we learned through our interviews, the level of abstraction seems to be lower than high-level source code. For example, engineers reported that they struggle to introduce abstraction such as nested scopes of variable visibility, enumerators, or refactoring duplicated code into a new method.

Compared to source-based high-level languages, we found that, while model-driven engineering increases the abstraction level of program compilation, it does achieve the same increase in abstraction for program representation. Model-driven engineering provides specialists with the power to build a domain-specific global architecture by customizing the program compilation through code-generation rules. Yet, the "end-users" of model-driven engineering, that is domain experts and software engineers, are left without the power to create their own APIs to address local problem-specific needs in a formal manner.

We are aware that our observations with regard to visual languages and the lack of domain-specific modeling are tied to the technology used in the setting under study, in particular Simulink's visual language and Rhapsody's use of UML. Our findings reflect the state of practice in one organization and are not necessarily representative of the latest state-of-the art in research or even other industries. In particular, domain-specific modeling (DSM) might alleviate the frictions discussed in this subsection [5]. In domain specific modeling the domain experts are empowered to specify the code generation such that modeling concepts map directly to domain concepts rather than computer technology concepts, thus overcoming the limitations of Simulink and UML.

## 6 Related Work

In this section we discuss related work, namely empirical studies of MDE. For a discussion of the state-of-the-art in, e.g., model diffing or other technologies related to the frictions presented in this paper, readers are advised to refer to recent proceedings of the MODELS conference and its workshops.

Although model-driven engineering claims many potential benefits, it has largely developed without the support of empirical data. There are few reports of empirical evaluations of modeling in the literature. Even fewer discuss human factors and cognitive issues of model-driven engineering, since most empirical studies has been focused technological aspect of MDE or UML in particular.

In parallel to our study, Aranda et.al. investigated the organizational consequences of adopting MDE at the same organization [6]. They interviewed the same participants as our second visit, but while we investigated cognitive issues of technology, driven largely from an individual's perspective, they looked into organizational forms, patterns, and processes of MDE adoption. They found that switching to MDE may disrupt organizational structure, creating morale and power problems. They conclude that the cultural and institutional infrastructure of MDE is underdeveloped and until MDE becomes better established, transitioning organizations need to exert additional adoption efforts.

Most recently Hutchinson et.al. presented their results of a qualitative user study, consisting of semi-structured interviews with 20 engineers in 20 different organizations [7,8]. They identified lessons learned, in particular the importance of complex organizational, managerial and social factors, as opposed to simple technical factors, in the relative success, or failure, of MDE. As an example of organizational change management, the successful deployment of model driven engineering appears to require: a progressive and iterative approach; transparent organizational commitment and motivation; integration with existing organizational processes and a clear business focus.

Mohagheghi and Dehlen presented a study on the impact of MDE on productivity and software quality [9]. Their methodology was a meta-analysis of the literature, selecting 25 papers published in quality conferences and venues between 2000 and 2007. Almost all these papers were experience reports from single projects and most of the papers present results anecdotally. Software processes were reported as being of integral importance in successfully applying MDE, and the importance of suitable tools was reported as of crucial importance. The meta-study also looked for evidence that MDE improves software quality, but the evidence was anecdotal. In conclusion, they suggested that there is a need for more empirical studies evaluating MDE before sufficient data will be available to prove the benefits of its use.

Forward and Lethbridge conducted a survey of 117 engineers to find practitioners' opinions and attitudes towards MDE [10]. In accordance with our findings, the study concludes that model-driven techniques may benefit from features that, synchronize code and models, better traceability between models and code, better modeling capabilities and expressibility the reduce the need for external artifacts. Alas, the survey provides little data on the participant's context, size of their organizations and their adoption-level of MDE. As it seems, only 32% of the participants reported that they generate all or some code from the models. Dobing and Parson discussed the survey to discover commonly-held perceptions which may not hold true in practice [11]. Anda et.al. reported on disadvantages of adopting modeling practices, such as the difficultly of integrating legacy code and models, but found anecdotal advantages of improved traceability [12]. Afonso et.al. wrote about a case study where developers migrate from code-centric to model-centric practices [13].

## 7  Conclusion

When technologies are introduced, it is often hard to separate myth from reality. To investigate the benefits and challenges of model-driven engineering we performed a field study about the use of model-based design in a large automotive company. We showed how, for one large organization, model-driven engineering is shaped by contextual forces, which seem to be independent of the abstraction chosen to help develop the system. Through this study, which involved interviews with 20 engineers and managers, we identified four forces and five points of friction (as itemized in the introduction). We differentiate between forces that are contextual and external to software modeling technologies and as frictions, which are accidental issues caused by current tooling on software modeling.

As we worked with the data, the contextual forces affecting individuals using modeling became clear. While architectural complexity is well hidden from software engineers, they are still exposed to substantial innate complexity (contextual forces) and often even new accidental complexity (points of frictions in modeling tools). In particular, the representational abstraction of visual modeling languages does not seem to be as broad and rich as the problem-specific visual and formal languages of the domain experts.

We identified three concluding themes that span across many of the identified forces and points of friction, which might be of interest for tool builder and language designers in their future work. They are as follows:

– Engineers seem to prefer the linear reading paths of textual representations over the spatial representation of nested visual models. Both when navigating and changing models as well as when using model diffing. They describe their experience as *"going blind"* and struggling *"to not miss anything."*
– While the MDE tools under study provide specialized code-generation engineers with powerful abstraction, they do not similarly empower its end-users, i.e., domain experts and software engineers, to express their own problem-specific languages and type systems.
– The needs of engineers who are inventing novel algorithms differ from those of engineers who are working on more mature features. Invention is an exploratory activity and its needs, such as instant model changes as runtime, seem not to be well addressed by current modeling tool-chains.

**Acknowledgments:** This research has been funded by the Canadian Network on Engineering Complex Software Intensive Systems for Automotive Applications (NECSIS). Special thanks to Joe D'Ambrosia and Shige Wang from GM Research, as well as Indira Nurdiani and Jorge Aranda, for their help conducting the interviews. We are grateful to all our study participants for their time and collaboration.